\documentclass[pra,twocolumn,showpacs,preprintnumbers,amsmath,amssymb]{revtex4-1}
\usepackage{graphicx}% Include figure files
\usepackage{dcolumn}% Align table columns on decimal point
\usepackage{bm}% bold math
\newcommand  {\comment}[1] { } % hide comments

\begin{document}

\title{Magnetic ordering of three-component ultracold fermionic mixtures in optical lattices}
%Magnetic Ordering of Three-Component Ultracold Fermionic Mixtures in Optical Lattices
\author{Andrii Sotnikov}
%\affiliation{Institut f\"ur Theoretische Physik, Goethe-Universit\"at, 60438 Frankfurt/Main, Germany}

\author{Walter Hofstetter}
\affiliation{Institut f\"ur Theoretische Physik, Goethe-Universit\"at, 60438 Frankfurt/Main, Germany}

\date{\today}% It is always \today, today,
             %  but any date may be explicitly specified

\begin{abstract}
We study finite-temperature magnetic phases of three-component mixtures of ultracold fermions with repulsive interactions in optical lattices with simple cubic or square geometry by means of dynamical mean-field theory (DMFT). We focus on the case of one particle per site (1/3 band filling) at moderate interaction strength, where we observe a sequence of thermal phase transitions into two- and three-sublattice ordered states by means of the unrestricted real-space generalization of DMFT. From our quantitative analysis we conclude that long-range ordering in three-component mixtures should be observable at comparable temperatures as in two-component mixtures.
\end{abstract}

% Lattice fermion models, 71.10.Fd
% Hubbard model
%  electronic structure, 71.10.Fd
%  magnetic ordering (quantized spin model), 75.10.Jm
% Ultracold gases, 67.85.-d
% Symmetry breaking, 11.30.Qc
% Symmetry in theory of fields and particles, 11.30.-j
% Optical cooling and trapping of atoms, 37.10.Jk
% 75.50.Ee -- antiferromagnetism in magnetic materials;
\maketitle

%% INTRODUCTION %%
\section{Introduction}
Multicomponent mixtures of ultracold fermions in optical lattices are novel and highly promising many-body systems that draw our attention to fundamental open questions such as collective excitations in the presence of high symmetries, color superfluidity, high-spin magnetism, and non-magnetic ground states as well as effects of breaking particular symmetries in the highly symmetric models. Due to the impressive experimental progress made in the past few years, a number of interesting phenomena peculiar to these systems have been already observed \cite{Tai2010PRL,Kra2012Nat,Tai2012Nat,Hei2013PRL,Kra2014S}. For the purpose of observation of ground-state magnetic properties in multicomponent mixtures, one requires additional cooling in these systems. In this direction, important progress was recently achieved towards understanding cooling mechanisms~\cite{BerPRA2009,Jor2010PRL,Lub2011PRL,Haz2012PRA} and observing short-range magnetic correlations \cite{Gre2013Sci} in optical lattices. In order 
to combine the progress in both directions and proceed further, quantitative predictions are needed concerning magnetic phases in ultracold mixtures of three and more components in optical lattices.

A number of theoretical studies have been performed in order to investigate ground-state phases in large-spin fermionic mixtures; in particular, on three-component mixtures in the context of color superfluidity and trionic phases \cite{Che2007PRL,Rap2008PRB,Tit2011NJP,Kan2012PRB}. For repulsive interactions, a multitude of magnetic ground states at different fillings has been predicted at zero temperature both in the case of full \cite{Rap2011PRA} and broken \cite{Ina2013MPL} SU(3) symmetry in the Hubbard Hamiltonian. At a filling of one particle per site (1/3 band filling) and strong interactions these mixtures can be described by the corresponding Heisenberg model. As it was argued in Ref.~\cite{Tot2010PRL}, the system described by an SU(3) Heisenberg model undergoes transitions between different sublattice orderings in simple lattice geometries at finite temperature.

As for higher-symmetry models, there is an ongoing debate about the ground states of the SU(4) Hubbard and corresponding Heisenberg models because of different and sometimes controversial predictions \cite{FAs2005PRB,Par2007JP,Cor2011PRL,Cai2013PRB}. As for SU($N\geq5$) models, most studies agree that long-range magnetic order is suppressed even at zero temperature \cite{Hon2004PRL,FAs2005PRB,Her2009PRL,Gor2010Nat,Sin2013PRA}.

In this paper, we focus on finite-temperature magnetic-ordering properties in three-component mixtures described by the Fermi-Hubbard Hamiltonian at moderately strong interactions.

\section{Model}\label{sec.2}
%{\it Model.---}
Since the physics of three-component fermi\-on\-ic mixtures is very rich, here we restrict ourselves to the tight-binding approximation valid for sufficiently strong optical lattices and low filling. In particular, we consider a Fermi-Hubbard Hamiltonian of the following type
\begin{eqnarray}
\mathcal{\hat{H}}=&&
-\sum\limits_{\langle ij\rangle}\sum\limits_{\alpha} t_\alpha \bigl( \hat{c}^\dag_{i\alpha}\hat{c}_{j\alpha}+{\rm H.c.}\bigr)
+\sum\limits_{i}\sum\limits_{\beta>\alpha}U_{\alpha\beta}\hat{n}_{i\alpha}\hat{n}_{i\beta}
\nonumber\\
&&+\sum\limits_{i}\sum\limits_{\alpha}(V_i-\mu_\alpha)\hat{n}_{i\alpha},
\label{eq.1}
\end{eqnarray}
where $t_\alpha$ is the hopping amplitude of fermionic species $\alpha=\{a,b,c\}$,
$\hat{c}^\dag_{i\alpha}$ ($\hat{c}_{i\alpha}$) is the corresponding creation (annihilation) operator of atoms $\alpha$ at the lattice site~$i$,
the notation $\langle ij\rangle$ indicates a summation over nearest-neighbor sites, and $U_{\alpha\beta}$ is the magnitude of the on-site repulsive ($U_{\alpha\beta}>0$ $\forall \alpha,\beta$) interaction of the two different species with corresponding densities $\hat{n}_{i\alpha}$ and $\hat{n}_{i\beta}$  ($\hat{n}_{i\alpha}=\hat{c}^\dag_{i\alpha}\hat{c}_{i\alpha}$). 
In the last term, $V_i$ is the external (e.g., harmonic) potential at lattice site $i$, and $\mu_\alpha$ is the chemical potential of species $\alpha$. 
Note that we have taken the harmonic potential to be independent of the atomic species.
The Hamiltonian~(\ref{eq.1}) implies a single-band approximation; in other words, we consider the case of a sufficiently strong lattice potential, $V_{\text{lat}}\gtrsim 5E_r$, where $E_r$ is the recoil energy of atoms. 

By using  the Schrieffer-Wolff transformation \cite{Mac1988PRB} in the limit $t_{\alpha}\ll U_{\alpha\beta}$ close to 1/3 band filling, $\sum_\alpha n_{i\alpha}\approx1$, one can map the Hamiltonian~(\ref{eq.1}) to an effective spin model.
For the system under study, this transformation results in
\begin{eqnarray}
 \mathcal{\hat{H}}_{\textrm{eff}} 
   =  
  -\sum\limits_{\langle ij\rangle}\sum\limits_{\beta>\alpha}
  \left(
  J^\parallel_{\alpha\beta} \hat{n}_{i\alpha}\hat{n}_{j\beta}
  -J^\perp_{\alpha\beta} \hat{c}^\dag_{i\alpha}\hat{c}_{i\beta}
  \hat{c}^\dag_{j\beta}\hat{c}_{j\alpha}
  \right)
  \nonumber 
  \\
  + (\mu_a-\mu_b)\sum_i\hat{S}_{3i}
  + \frac{1}{\sqrt3}(\mu_a+\mu_b-2\mu_c)\sum_i\hat{S}_{8i}\,,
  \label{eq.2}
\end{eqnarray}
where $\hat{S}_{k}$ is the pseudospin projection operator [the generator of the SU(3) group] to the $k$th axis in the effective eight-dimensional spin space. It can be expressed through Gell-Mann matrices \cite{Georgi1999} $\boldsymbol{\lambda}_k=\{\boldsymbol{\lambda}_1,\ldots,\boldsymbol{\lambda}_8 \}$ in a way analogous to the spin-1/2 case, $\hat{S}_{k} = \frac{1}{2} \hat{c}^\dag_{\alpha} \lambda_{k\alpha\beta} \hat{c}_{\beta}$ (see Appendix~\ref{A1} for details). The effective magnetic couplings are $J^\parallel_{\alpha\beta}=2(t_\alpha^2+t_\beta^2)/U_{\alpha\beta}$ and $J^\perp_{\alpha\beta}=4t_\alpha t_\beta/U_{\alpha\beta}$. Analogously to the spin-1/2 XXZ model, these couplings determine the relative strength of the easy-axis (Ising-type, $J^\parallel$) or easy-plane (XY type, $J^\perp$) magnetic correlations that influence the choice of the corresponding magnetic ground states in the system (see Appendices~\ref{A2} and \ref{A3} for details).

\section{Method}\label{sec.3}
%{\it Method.---}
We use dynamical mean-field theory (DMFT) \cite{Geo1996RMP} and its real-space generalization (R-DMFT) \cite{Hel2008PRL,Sno2008NJP} to study magnetic ordering properties in the system under consideration.  DMFT maps the lattice problem to an impurity problem, thus substituting the full action with an effective one. Despite the fact that it is a nonperturbative approach, it is still an approximate method, since it treats the lattice self-energy as a local quantity, thus neglecting nonlocal quantum fluctuations. DMFT becomes exact in the limit of infinite spatial dimensionality, $d=\infty$ (i.e., large coordination number $z\gg1$). Although it is not an exact method in the case of square and cubic lattice geometries ($z=4$ and $z=6$, respectively), results obtained by DMFT are a reference point both for experiments and for more sophisticated methods, such as quantum Monte Carlo simulations, which are computationally rather demanding due to the generic presence of a sign problem for 
fermionic mixtures with an odd number of spin components \cite{Wu2005PRB}.

For solving the effective quantum impurity problem  we choose the continuous-time Monte-Carlo hybridization expansion solver (CT-HYB) \cite{Gul2011RMP,Buchhold2012}, since it allows a rather straightforward generalization to the case of an arbitrary number of spin components in the system. Here we restrict ourselves to measuring observables that are diagonal in spin space, but the approach can be generalized to off-diagonal quantities as well.

Since we are interested in predictions for experimentally relevant lattice geometries (e.g., square or cubic), after solving the impurity problem it is necessary to express all relevant quantities in Matsubara-frequency space (this procedure can be avoided only for the infinite-dimensional Bethe lattice due to its loopless structure \cite{Comanac2007}). To reduce the noise in the self-energy, we use a Legendre-polynomial representation \cite{Boe2011PRB} and calculate both two- and four-point Green's functions \cite{Haf2012PRB}. Then, one can generalize the expression for the self-energy \cite{Bulla1998JP,Haf2012PRB} and write it as follows
\begin{equation}
 \Sigma_\alpha(i\omega_n) = \sum_{\beta>\alpha}U_{\alpha\beta} {F}_{\alpha\beta}(i\omega_n)/{G}_\alpha(i\omega_n),
\end{equation}
where ${F}_{\alpha\beta}(i\omega_n)$ is the Fourier transform of the corresponding four-point Green's function ${F}_{\alpha\beta}(\tau-\tau')=-\langle{\cal T}_\tau \hat{c}_\alpha(\tau)\hat{c}^\dag_\alpha(\tau')\hat{n}_\beta (\tau')\rangle$.

Within R-DMFT, the obtained self-energies are input into the real-space matrix (lattice Dyson equation)
\begin{equation}
 ({\bf G}^{-1}_\alpha)_{ij} = (i\omega_n + \mu_\alpha - V_i - \Sigma_{\alpha i})\delta_{ij}+{t}_{\alpha ij}\,,
\label{eq.rdmft}
\end{equation}
where ${t}_{\alpha ij}$ is the hopping matrix element ($i$ and $j$ are site indices).
Inversion of Eq.~(\ref{eq.rdmft}) results in a matrix containing the local lattice Green's functions along the main diagonal.
Then, we use the local Dyson equation 
%(the repeated lattice indices are omitted below)
\begin{equation}
 {\cal G}_{\alpha i}^{-1}(i\omega_n) = G_{\alpha ii}^{-1}(i\omega_n) + \Sigma_{\alpha i}(i\omega_n),
\end{equation}
where ${\cal G}_{\alpha i}$ is the Weiss function (i.e., the dynamical mean field) of the effective impurity model at site~$i$,
and express the corresponding hybridization function for the new DMFT iteration
\begin{equation}
 \Gamma_{\alpha i}(-i\omega_n) = i\omega_n + \mu_\alpha - {\cal G}_{\alpha i}^{-1}(i\omega_n).
\label{eq.6}
\end{equation}
To reduce the error in the new hybridization function on the imaginary-time axis $\Gamma(\tau)$ originating from the numerical inverse Fourier transformation (IFT), we use the knowledge about its asymptotic behavior at large frequencies \cite{Comanac2007}
\begin{equation}
\left[\Gamma_{\alpha i} (-i\omega_n) - \frac{\langle\epsilon^2\rangle}{i\omega_n}\right]
\quad
\underrightarrow{\;\mbox{IFT}\;}%\longrightarrow
\quad
\left[\Gamma_{\alpha i} (\tau) + \frac{\langle\epsilon^2\rangle}{2} \right].
\end{equation}
The quantity $\langle\epsilon^2\rangle$ can be obtained by using the non-interacting density of states $D(\epsilon)$ for the considered lattice geometry, $\langle\epsilon^2\rangle = \int D(\epsilon)\epsilon^2d\epsilon$.

\section{Results}\label{sec.4}
\begin{figure}
\includegraphics[width=\linewidth]{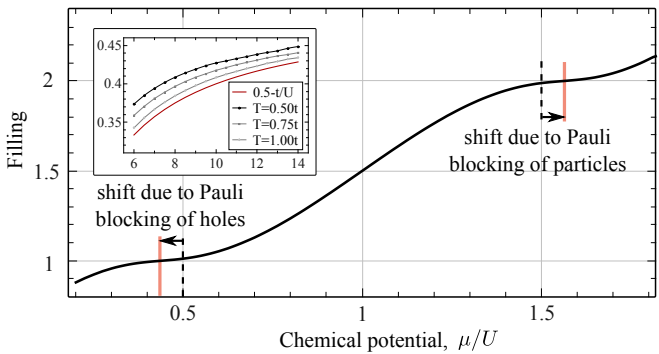}
%Fig.1
    \caption{\label{Pauli} (color online)
    Dependence of the filling per site on the chemical potential for an SU(3)-symmetric mixture in the cubic lattice at $U=10t$ and $T=0.5t$ (paramagnetic region). (Inset) Dependence of the chemical potential $\mu/U$ on the interaction strength $U/t$ at filling $n=1$ for different temperatures, which shows that the shift is proportional to $t/U$.}
\end{figure}

%{\it Results.---}
In our studies we focus on the case of 1/3 band filling. It should be mentioned that there is no exact expression for the chemical potential value as for the case of half filling. For the SU($N$)-symmetric mixture at half filling one can derive a general relation $\mu = \frac{U}{2}(N-1)$, which is due to the particle-hole symmetry in the Hubbard Hamiltonian. Evidently, for other fillings there is no such symmetry, and effects originating from Pauli blocking can have a strong impact. In particular, from Fig.~\ref{Pauli} we conclude that the approximate condition $\mu = U/2$ guarantees occupation of one particle per site only deep inside the insulator region. At weak and moderate interaction strength, the chemical potentials must be additionally adjusted in order to have a proper filling in the system.

\begin{figure}
\includegraphics[width=\linewidth]{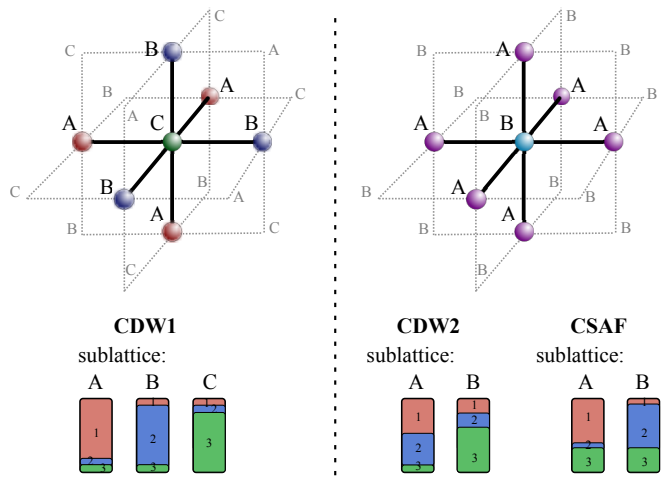}
% Fig.2
    \caption{\label{su3phases} (color online)
    Sketch of two types of sublattice ordering (above) and the corresponding three types of magnetic phases (below) observed in our R-DMFT analysis at 1/3 band filling for a three-component fermionic mixture. The depicted main two-sublattice ordered states (right) can be understood in the following way: CDW2 -- ``1'' and ``2'' team up against ``3'',  CSAF -- ``3'' is left out of AF correlations.}
\end{figure}

To resolve different sublattice orderings in a three-component mixture at low temperatures, as was argued in Ref.~\cite{Tot2010PRL} for the Heisenberg model, we apply the R-DMFT method for the system described by Eq.~(\ref{eq.1}). 
%%% CHANGE
We use periodic boundary conditions, which enter our numerical analysis through the matrix elements~$t_{\alpha ij}$ in Eq.~(\ref{eq.rdmft}), and system sizes with numbers of sites in each direction that are integer multiples of spatial periods of the ordering structure in the system, e.g. $6^3$ and $12^3$ sites.
%%%
Depending on the Hubbard parameters and temperature,
%%%
 we observe several magnetic ground states which are sketched in Fig.~\ref{su3phases} (here and below we use numbers for the spin indices). We identify three-sublattice order corresponding to the color density wave (CDW1)
%%% CHANGE
 with the ordering wave vector ${\bf Q}=(\pm2\pi/3,\pm2\pi/3,\pm2\pi/3)$ (all $\pm$ signs are independent of each other)
%%%
 and two-sublattice order
%%% CHANGE
 characterized by ${\bf Q}=(\pi,\pi,\pi)$
%%%
 with two distinct antiferromagnetic states: a second type of color density wave (CDW2) or color-selective antiferromagnet (CSAF).

\begin{figure}
\includegraphics[width=\linewidth]{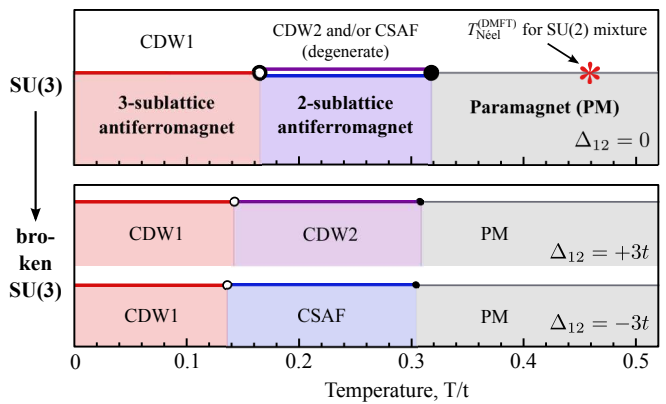}
% Fig.3
    \caption{\label{su3not} (color online)
    Transitions between sublattice orderings at finite temperature obtained by R-DMFT for a cubic lattice ($12^3$ sites). Parameters are $U_{12}={U}+\Delta_{12}$, $U_{13}=U_{23}={U}-\Delta_{12}/2$, $U=12t$.}
\end{figure}

To study transitions between these states, we perform calculations at different temperatures.
%%% CHANGE
We analyze the hybridization functions~(\ref{eq.6}) on each lattice site~$i$ for every DMFT iteration and in case of successful convergence of the program we collect output data from the impurity solver, in particular, expectation values of the number operator $\langle\hat{n}_{i\alpha}\rangle$ for each spin component $\alpha$. The spatial periodicity and values of these quantities allow us to directly identify the phases sketched in Fig.~\ref{su3phases} and the corresponding critical temperatures.
%%%
In Fig.~\ref{su3not} we show the observed magnetic phases both in the presence of the full SU(3) symmetry and in the case of broken symmetry due to different interspecies interactions.

From Fig.~\ref{su3not} one can analyze critical temperatures for different sublattice orderings. It shows, in particular, that the ordering phenomenon should be observed at moderate temperatures of the order of the superexchange amplitude $t^2/U$ as in the spin-1/2 case. As we see, the two-sublattice phases (CDW2 and CSAF) are preferred by thermal fluctuations, while three-sublattice order (CDW1) is preferred by quantum fluctuations, which is in perfect agreement with the reasoning in \cite{Tot2010PRL} based on a semiclassical analysis.
%%% CHANGE
Note that we also performed the R-DMFT analysis for the square lattice, which shows the same structure of the phases as shown in Fig.~\ref{su3not} with lower numerical values for all critical temperatures due to the lower coordination number of the square lattice.
%%%

We can also make some important statements about magnetic phases in the presence of different interspecies interactions that can be realized in alkali atoms by means of Feshbach resonances. First, it should be mentioned that this asymmetry in interaction removes the degeneracy in the two-sublattice ground states in the same way as in the case of half filling studied in Ref.~\cite{Ina2013MPL} (here, the choice of the particular ground state is determined by the ratio between different magnetic couplings $J^\parallel_{\alpha\beta}$). Second, the observed suppression of critical temperatures for magnetic ordering can be explained from the analysis of the effective spin model described by Eq.~(\ref{eq.2}). Note that the CDW1 order involves all three magnetic couplings $J^\parallel_{\alpha\beta}$, thus the critical temperature strongly depends on the minimal magnetic coupling, which in the asymmetric cases shown in Fig.~\ref{su3not} is lower than in the $SU(3)$-symmetric case. 
The additional suppression of the two-sublattice phases in the asymmetric regimes can be explained by energetic arguments based on the analysis of the term $\mathcal{\hat{H}}_U= \sum_{i}\sum_{\beta>\alpha}U_{\alpha\beta} \hat{n}_{i\alpha}\hat{n}_{i\beta}$ in Eq.~(\ref{eq.1}).
%%% CHANGE
In fact, the change of the interaction asymmetry by tuning only the parameter $\Delta_{12}$ as in Fig.~\ref{su3not} leaves the eigenvalues of $\mathcal{\hat{H}}_U$ unchanged in the paramagnetic region independently of the $\Delta_{12}$ value (assuming that all the resulting $U_{\alpha\beta}$ remain positive). But in the ordered region (CDW2 or CSAF), the eigenvalues of $\mathcal{\hat{H}}_U$ are different for different values of $\Delta_{12}$. In particular, the state CDW2 becomes energetically penalized with an increase of $\Delta_{12}>0$ and the state CSAF becomes less energetically favorable with a decrease of $\Delta_{12}<0$.
%%%

\begin{figure}
\includegraphics[width=\linewidth]{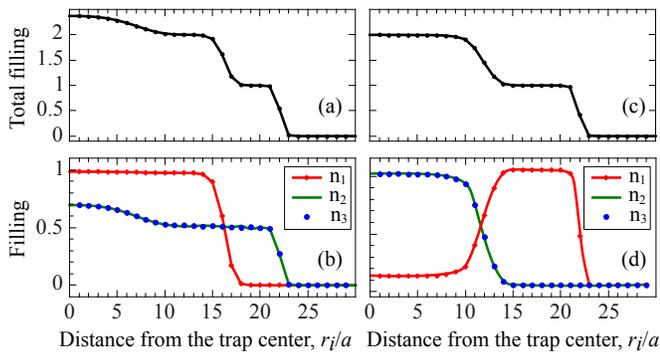}
    \caption{\label{separation} (color online)
    Real-space total density and spin density profiles obtained by DMFT in a cubic lattice for anisotropic interaction strengths. Parameters are $T=4t$, $V_i=0.04t(r_i/a)^2$, where $a$ is the lattice spacing;
    (a)--(b): $U_{12}=U_{13}=20t$, $U_{23}=40t$, $\mu_1=45t$, $\mu_2=\mu_3=60t$; 
    (c)--(d): $U_{12}=U_{13}=40t$, $U_{23}=20t$, $\mu_1=60t$, $\mu_2=\mu_3=45t$.}
\end{figure}

Here, we note one more experimentally relevant effect originating from the asymmetry in the interaction strengths $U_{\alpha\beta}$. The above-mentioned breaking of SU(3) symmetry by means of Feshbach resonances can lead to a strong separation of spin components in a harmonic trap. In Fig.~\ref{separation} we show this effect for a range of temperatures and asymmetries higher than considered above.
%%% CHANGE
This choice is motivated by the current experimental possibilities for the temperatures achievable in ultracold fermionic mixtures. Furthermore, it allows us to exclude any effects based on the superexchange processes. Naturally, the mentioned spin separation also takes place at lower temperatures with an even more pronounced separation effect.

%%%
The asymmetry in the interaction strengths effectively results in an additional spin-dependent trapping potential. Due to this mechanism, colorful Mott-shell structures consisting of a single or multiple components arise. These structures are also interesting with respect to spin-dependent dynamical properties and possible applications to many-body cooling, since they contain significantly less entropy per particle than in the SU(3)-symmetric case \cite{BerPRA2009,Lub2011PRL}. The latter fact, in particular, could help in experiments to reach and observe the ground-state magnetic phases described above.

\section{Conclusions}
%{\it Conclusions.---}
We studied finite-temperature properties of magnetic long-range order in three-component mixtures of ultracold fermions with repulsive interactions in optical lattices by means of real-space dynamical mean-field theory.
We showed that at 1/3 band filling with increasing temperature the system undergoes a sequence of thermal phase transitions between different sublattice orderings, which agrees in the limit $U\gg t$ with previous predictions for the SU(3) Heisenberg model.
We also studied magnetic ordering in three-component mixtures with different interspecies interactions that can be realized in alkali atoms by means of Feshbach resonances. It is shown that an asymmetric interaction of this type removes degeneracy in two-sublattice orderings and leads to a suppression of critical temperatures for both two- and three-sublattice ordering.

\begin{acknowledgments}
The authors thank M.~Buchhold, D.~Cocks and J.~Krauser for useful discussions.
Support by the German Science Foundation DFG via Sonderforschungsbereich SFB/TR 49 and Forschergruppe FOR 801 is gratefully acknowledged.
\end{acknowledgments}

\begin{appendix}

\section{Gell-Mann matrices and SU(3) generators}\label{A1}

The Gell-Mann matrices introduced in the main part of the paper are defined as follows:
\begin{eqnarray*}
 &&\boldsymbol{\lambda}_1=\left(\begin{array}{c c c}
 0 & 1  & 0\\
 1 & 0  & 0\\
 0 & 0  & 0\end{array}\right),
~
 \boldsymbol{\lambda}_2=\left(\begin{array}{c c c}
 0 & -i & 0\\
 i & 0  & 0\\
 0 & 0  & 0\end{array}\right),
\\
 &&\boldsymbol{\lambda}_4=\left(\begin{array}{c c c}
 0 & 0  & 1\\
 0 & 0  & 0\\
 1 & 0  & 0\end{array}\right),
~
 \boldsymbol{\lambda}_5=\left(\begin{array}{c c c}
 0 & 0 & -i\\
 0 & 0  & 0\\
 i & 0  & 0\end{array}\right),
\\
 &&\boldsymbol{\lambda}_6=\left(\begin{array}{c c c}
 0 & 0  & 0\\
 0 & 0  & 1\\
 0 & 1  & 0\end{array}\right),
~
 \boldsymbol{\lambda}_7=\left(\begin{array}{c c c}
 0 & 0  & 0\\
 0 & 0 & -i\\
 0 & i  & 0\end{array}\right),
\\
&&
 \boldsymbol{\lambda}_3=\left(\begin{array}{c c c}
 1 & 0  & 0\\
 0 & -1 & 0\\
 0 & 0  & 0\end{array}\right),
~
 \boldsymbol{\lambda}_8=\frac{1}{\sqrt3}\left(\begin{array}{c c c}
 1 & 0  & 0\\
 0 & 1  & 0\\
 0 & 0 & -2\end{array}\right).
\end{eqnarray*}
Therefore, the introduced pseudospin operators $\hat{S}_{k} = \frac{1}{2} \hat{c}^\dag_{\alpha} \lambda_{k\alpha\beta} \hat{c}_{\beta}$ can be considered as the generators of the SU(3) group, since the operator
\begin{equation*}
 \hat{\cal U} = \exp\left( i \sum_{k=1}^8\varphi_k \hat{S}_{k} \right)
\end{equation*}
performs special ($\det {\cal U}=1$) unitary rotations in the corresponding space ($\varphi_k$ are real numbers).

The pseudospin operators obey the commutation relations
\begin{equation*}
 [\hat{S}_{i},\hat{S}_{j}] = i\sum_{k=1}^8f^{ijk}\hat{S}_{k},
\end{equation*}
where the structure constants $f^{ijk}$ are completely antisymmetric in the three indices and have values
\begin{eqnarray*}
 && f^{123}=1,\qquad f^{458}=f^{678}={\sqrt{3}}/{2},\\
 && f^{147}=f^{165}=f^{246}=f^{257}=f^{345}=f^{376}={1}/{2},
\end{eqnarray*}
while all other $f^{ijk}$ not related to these by permutation are zero.

\section{Effective pseudospin Hamiltonian}\label{A2}
 The effective pseudospin Hamiltonian given by Eq.~(2) provides important information about possible ground-state magnetic phases in the system. For example, from the first term with $J^\parallel_{\alpha\beta}$ we note that, in order to minimize the energy, the system prefers to have nearest-neighbor sites occupied by different components (this corresponds to easy-axis antiferromagnetic ordering). For completeness, here we show the form of this Hamiltonian solely in terms of the pseudospin operators $\hat{S}_{k}$.
 
 Using the expressions for the number operators at filling $n\equiv\sum_\alpha n_{\alpha}=1$ (1/3 band filling) at given lattice site~$i$ ($a,b,c$ are the spin indices)
\begin{eqnarray*}
 && \hat{n}_a + \hat{n}_b = \frac{2}{3}\hat{n}+\frac{2}{\sqrt{3}}\hat{S}_{8},
    \quad\qquad~
   \hat{n}_a - \hat{n}_b = 2\hat{S}_{3},
 \\
 && \hat{n}_a + \hat{n}_c = \frac{2}{3}\hat{n}+\hat{S}_{3}-\frac{1}{\sqrt{3}}\hat{S}_{8},
    \quad
   \hat{n}_a - \hat{n}_c = \sqrt{3}\hat{S}_{8}+\hat{S}_{3},
  \\
 && \hat{n}_b + \hat{n}_c = \frac{2}{3}\hat{n}-\hat{S}_{3}-\frac{1}{\sqrt{3}}\hat{S}_{8},
    \quad\,
   \hat{n}_b - \hat{n}_c = \sqrt{3}\hat{S}_{8}-\hat{S}_{3},
\end{eqnarray*} 
we can write Eq.~(2) in the form
\begin{eqnarray}
  \mathcal{\hat{H}}_{\textrm{eff}} 
   ~\quad\qquad\qquad\qquad\qquad\qquad\qquad\qquad\qquad\qquad\qquad\quad\
  \nonumber
\\
  =\sum\limits_{\langle ij\rangle}
  \left[
    J^\parallel_{ab}\left(
    \hat{S}_{3i}\hat{S}_{3j}-\frac{1}{3}\hat{S}_{8i}\hat{S}_{8j}
    \right)
  +J^\perp_{ab}\left(\hat{S}_{1i}\hat{S}_{1j}+\hat{S}_{2i}\hat{S}_{2j}\right)
  \right.
  \nonumber
\\
    +J^\parallel_{ac}\left(
    \frac{2}{3}\hat{S}_{8i}\hat{S}_{8j}+\frac{2}{\sqrt{3}}\hat{S}_{3i}\hat{S}_{8j}
    \right)
  +J^\perp_{ac}\left(\hat{S}_{4i}\hat{S}_{4j}+\hat{S}_{5i}\hat{S}_{5j}\right)
  \nonumber
\\
  \left.
    +J^\parallel_{bc}\left(
    \frac{2}{3}\hat{S}_{8i}\hat{S}_{8j}-\frac{2}{\sqrt{3}}\hat{S}_{3i}\hat{S}_{8j}
    \right)
  +J^\perp_{bc}\left(\hat{S}_{6i}\hat{S}_{6j}+\hat{S}_{7i}\hat{S}_{7j}\right)
  \right]
  \nonumber
\\
  + \left[(\mu_a-\mu_b)-(J^\parallel_{ac}
    -J^\parallel_{bc})/3\right]\sum_i\hat{S}_{3i}
  \nonumber
\\
  + \frac{1}{\sqrt3}\left[(\mu_a+\mu_b-2\mu_c)-(2J^\parallel_{ab}
      -J^\parallel_{ac}-J^\parallel_{bc})/3
    \right]\sum_i\hat{S}_{8i}\,,
  \nonumber
\\
\label{eq.A1}
\end{eqnarray}
where $J^\parallel_{\alpha\beta}=2(t_\alpha^2+t_\beta^2)/U_{\alpha\beta}$ and $J^\perp_{\alpha\beta}=4t_\alpha t_\beta/U_{\alpha\beta}$.
From here we note that the asymmetry in interactions can additionally induce the effective magnetic field in the system under study. This situation differs from two-component mixtures, where the effective magnetic field can be induced only by a finite chemical potential difference. In numerical calculations for asymmetric cases, in order to have balanced average occupation by each spin component ($\bar{n}_{i\alpha}=1/3$ $\forall\alpha$), we compensate the magnetic field induced in this way by the corresponding difference in chemical potentials.

In the case of complete SU(3) symmetry ($t_\alpha=t$, $U_{\alpha\beta}=U$, $\mu_\alpha=\mu$ $\forall\alpha,\beta$) it is easy to verify that the above Hamiltonian reduces to the SU(3) Heisenberg model
\begin{eqnarray}
  \mathcal{\hat{H}}_{\textrm{eff}} = J
  \sum\limits_{\langle ij\rangle}
  \sum\limits_{k=1}^{8} \hat{S}_{ki}\hat{S}_{kj}
\label{eq.A2}
\end{eqnarray}
with positive (antiferromagnetic) exchange coupling $J=4t^2/U$.

\section{Experimental possibilities and symmetry breakings in the Hamiltonian}\label{A3}

From the effective Hamiltonians~(2), (\ref{eq.A1}), and (\ref{eq.A2}) we can draw several important conclusions about symmetries that are present or can be broken if one introduces different interspecies interactions, imbalances in hopping amplitudes, or chemical potentials.

In the case of complete SU(3) symmetry one arrives at the Heisenberg model without anisotropies in the magnetic couplings or effective magnetic fields induced by differences in chemical potentials. The spontaneous symmetry breaking, which corresponds to the transition to a magnetically-ordered (antiferromagnetic) state, gives rise to six Goldstone modes in the system. With the three-component mixtures of ultracold fermions in optical lattices one can reduce the SU(3) symmetry in the Hamiltonian (thus suppressing the number of gapless excitations) by three different mechanisms that can be applied independently or in a combination, if necessary.

First, if one introduces different chemical potentials $\mu_{\alpha}\neq\mu_{\beta}~\forall{\alpha,\beta}$, but keeps all other parameters symmetric, the system may prefer a canted configuration: the net (non-staggered) magnetizations point along the directions $\boldsymbol{\lambda}_3$ and $\boldsymbol{\lambda}_8$. The easy-axis antiferromagnetic correlations [described by the terms $J^\parallel_{\alpha\beta}$ in Eqs.~(2) and (\ref{eq.A1})] are thus suppressed and the system develops antiferromagnetic order along other quantization axes. There are in general two gapless excitations [two continuous symmetries $U(1)\times U(1)$ generated by $\hat{S}_{3}$ and $\hat{S}_{8}$] in this state. Note that in the special case when one effective magnetic field is zero, e.g., $\mu_{a}=\mu_{b}\neq\mu_{c}$, there are four continuous symmetries [the corresponding group $SU(2)\times U(1)$ generated by $\hat{S}_{1}$, $\hat{S}_{2}$, $\hat{S}_{3}$, and $\hat{S}_{8}$] in the Hamiltonian.

Second, the initial symmetry of the model can be broken by different hopping amplitudes $t_{\alpha}\neq t_{\beta}~\forall{\alpha,\beta}$ (``complete mass imbalance''). This reduces the symmetry of the Hamiltonian in the same way as different chemical potentials, leaving two (or four in the case of $t_{\alpha}=t_{\beta}\neq t_{\gamma}$) continuous symmetries. Here, the important difference is in the ground states, since in the case of absence or sufficiently weak effective magnetic fields the system prefers easy-axis (``natural color'') antiferromagnetic order (note that $J^\parallel_{\alpha\beta}> J^\perp_{\alpha\beta}$ due to mass imbalance and the interaction strengths are assumed symmetric, $U_{\alpha,\beta}=U~\forall{\alpha,\beta}$). These states may have significant advantages at finite temperatures \cite{Sotnikov2012PRL,Sotnikov2013PRA}, since in the case of complete mass imbalance they have no gapless long-wave excitations.

Third, the interaction strengths $U_{\alpha\beta}$ can be tuned by Feshbach resonances or (and) using mixtures of different atoms, isotopes, or metastable excited states of the same atom. The symmetry breaking caused by asymmetric interactions reduces the initial symmetry of the Hamiltonian to the same continuous groups as in the above cases (here we consider $\mu_{\alpha}=\mu_{\beta}~\forall{\alpha,\beta}$). Note that the asymmetry in interactions generates additional effective magnetic fields [see Eq.~(\ref{eq.A1})]. Therefore, to make the easy-axis antiferromagnetic correlations dominating (in the case $J^\parallel_{\alpha\beta}> J^\perp_{\alpha\beta}$) or, at least, the same (in the case $J^\parallel_{\alpha\beta}= J^\perp_{\alpha\beta}$), these fields must be compensated by the corresponding differences in chemical potentials.
 
\end{appendix}

\bibliography{A16}% Produces the bibliography via BibTeX.

%merlin.mbs apsrev4-1.bst 2010-07-25 4.21a (PWD, AO, DPC) hacked
%Control: key (0)
%Control: author (8) initials jnrlst
%Control: editor formatted (1) identically to author
%Control: production of article title (-1) disabled
%Control: page (0) single
%Control: year (1) truncated
%Control: production of eprint (0) enabled
\begin{thebibliography}{39}%
\makeatletter
\providecommand \@ifxundefined [1]{%
 \@ifx{#1\undefined}
}%
\providecommand \@ifnum [1]{%
 \ifnum #1\expandafter \@firstoftwo
 \else \expandafter \@secondoftwo
 \fi
}%
\providecommand \@ifx [1]{%
 \ifx #1\expandafter \@firstoftwo
 \else \expandafter \@secondoftwo
 \fi
}%
\providecommand \natexlab [1]{#1}%
\providecommand \enquote  [1]{``#1''}%
\providecommand \bibnamefont  [1]{#1}%
\providecommand \bibfnamefont [1]{#1}%
\providecommand \citenamefont [1]{#1}%
\providecommand \href@noop [0]{\@secondoftwo}%
\providecommand \href [0]{\begingroup \@sanitize@url \@href}%
\providecommand \@href[1]{\@@startlink{#1}\@@href}%
\providecommand \@@href[1]{\endgroup#1\@@endlink}%
\providecommand \@sanitize@url [0]{\catcode `\\12\catcode `\$12\catcode
  `\&12\catcode `\#12\catcode `\^12\catcode `\_12\catcode `\%12\relax}%
\providecommand \@@startlink[1]{}%
\providecommand \@@endlink[0]{}%
\providecommand \url  [0]{\begingroup\@sanitize@url \@url }%
\providecommand \@url [1]{\endgroup\@href {#1}{\urlprefix }}%
\providecommand \urlprefix  [0]{URL }%
\providecommand \Eprint [0]{\href }%
\providecommand \doibase [0]{http://dx.doi.org/}%
\providecommand \selectlanguage [0]{\@gobble}%
\providecommand \bibinfo  [0]{\@secondoftwo}%
\providecommand \bibfield  [0]{\@secondoftwo}%
\providecommand \translation [1]{[#1]}%
\providecommand \BibitemOpen [0]{}%
\providecommand \bibitemStop [0]{}%
\providecommand \bibitemNoStop [0]{.\EOS\space}%
\providecommand \EOS [0]{\spacefactor3000\relax}%
\providecommand \BibitemShut  [1]{\csname bibitem#1\endcsname}%
\let\auto@bib@innerbib\@empty
%</preamble>
\bibitem [{\citenamefont {Taie}\ \emph {et~al.}(2010)\citenamefont {Taie},
  \citenamefont {Takasu}, \citenamefont {Sugawa}, \citenamefont {Yamazaki},
  \citenamefont {Tsujimoto}, \citenamefont {Murakami},\ and\ \citenamefont
  {Takahashi}}]{Tai2010PRL}%
  \BibitemOpen
  \bibfield  {author} {\bibinfo {author} {\bibfnamefont {S.}~\bibnamefont
  {Taie}}, \bibinfo {author} {\bibfnamefont {Y.}~\bibnamefont {Takasu}},
  \bibinfo {author} {\bibfnamefont {S.}~\bibnamefont {Sugawa}}, \bibinfo
  {author} {\bibfnamefont {R.}~\bibnamefont {Yamazaki}}, \bibinfo {author}
  {\bibfnamefont {T.}~\bibnamefont {Tsujimoto}}, \bibinfo {author}
  {\bibfnamefont {R.}~\bibnamefont {Murakami}}, \ and\ \bibinfo {author}
  {\bibfnamefont {Y.}~\bibnamefont {Takahashi}},\ }\href {\doibase
  10.1103/PhysRevLett.105.190401} {\bibfield  {journal} {\bibinfo  {journal}
  {Phys. Rev. Lett.}\ }\textbf {\bibinfo {volume} {105}},\ \bibinfo {pages}
  {190401} (\bibinfo {year} {2010})}\BibitemShut {NoStop}%
\bibitem [{\citenamefont {Krauser}\ \emph {et~al.}(2012)\citenamefont
  {Krauser}, \citenamefont {Heinze}, \citenamefont {Fl\"aschner}, \citenamefont
  {G\"otze}, \citenamefont {J\"urgensen}, \citenamefont {L\"uhmann},
  \citenamefont {Becker},\ and\ \citenamefont {Sengstock}}]{Kra2012Nat}%
  \BibitemOpen
  \bibfield  {author} {\bibinfo {author} {\bibfnamefont {J.~S.}\ \bibnamefont
  {Krauser}}, \bibinfo {author} {\bibfnamefont {J.}~\bibnamefont {Heinze}},
  \bibinfo {author} {\bibfnamefont {N.}~\bibnamefont {Fl\"aschner}}, \bibinfo
  {author} {\bibfnamefont {S.}~\bibnamefont {G\"otze}}, \bibinfo {author}
  {\bibfnamefont {O.}~\bibnamefont {J\"urgensen}}, \bibinfo {author}
  {\bibfnamefont {D.-S.}\ \bibnamefont {L\"uhmann}}, \bibinfo {author}
  {\bibfnamefont {C.}~\bibnamefont {Becker}}, \ and\ \bibinfo {author}
  {\bibfnamefont {K.}~\bibnamefont {Sengstock}},\ }\href {\doibase
  10.1038/nphys2409} {\bibfield  {journal} {\bibinfo  {journal} {Nat. Phys.}\
  }\textbf {\bibinfo {volume} {8}},\ \bibinfo {pages} {813} (\bibinfo {year}
  {2012})}\BibitemShut {NoStop}%
\bibitem [{\citenamefont {Taie}\ \emph {et~al.}(2012)\citenamefont {Taie},
  \citenamefont {Yamazaki}, \citenamefont {Sugawa},\ and\ \citenamefont
  {Takahashi}}]{Tai2012Nat}%
  \BibitemOpen
  \bibfield  {author} {\bibinfo {author} {\bibfnamefont {S.}~\bibnamefont
  {Taie}}, \bibinfo {author} {\bibfnamefont {R.}~\bibnamefont {Yamazaki}},
  \bibinfo {author} {\bibfnamefont {S.}~\bibnamefont {Sugawa}}, \ and\ \bibinfo
  {author} {\bibfnamefont {Y.}~\bibnamefont {Takahashi}},\ }\href {\doibase
  10.1038/nphys2430} {\bibfield  {journal} {\bibinfo  {journal} {Nat. Phys.}\
  }\textbf {\bibinfo {volume} {8}},\ \bibinfo {pages} {825} (\bibinfo {year}
  {2012})}\BibitemShut {NoStop}%
\bibitem [{\citenamefont {Heinze}\ \emph {et~al.}(2013)\citenamefont {Heinze},
  \citenamefont {Krauser}, \citenamefont {Fl\"aschner}, \citenamefont
  {Sengstock}, \citenamefont {Becker}, \citenamefont {Ebling}, \citenamefont
  {Eckardt},\ and\ \citenamefont {Lewenstein}}]{Hei2013PRL}%
  \BibitemOpen
  \bibfield  {author} {\bibinfo {author} {\bibfnamefont {J.}~\bibnamefont
  {Heinze}}, \bibinfo {author} {\bibfnamefont {J.~S.}\ \bibnamefont {Krauser}},
  \bibinfo {author} {\bibfnamefont {N.}~\bibnamefont {Fl\"aschner}}, \bibinfo
  {author} {\bibfnamefont {K.}~\bibnamefont {Sengstock}}, \bibinfo {author}
  {\bibfnamefont {C.}~\bibnamefont {Becker}}, \bibinfo {author} {\bibfnamefont
  {U.}~\bibnamefont {Ebling}}, \bibinfo {author} {\bibfnamefont
  {A.}~\bibnamefont {Eckardt}}, \ and\ \bibinfo {author} {\bibfnamefont
  {M.}~\bibnamefont {Lewenstein}},\ }\href {\doibase
  10.1103/PhysRevLett.110.250402} {\bibfield  {journal} {\bibinfo  {journal}
  {Phys. Rev. Lett.}\ }\textbf {\bibinfo {volume} {110}},\ \bibinfo {pages}
  {250402} (\bibinfo {year} {2013})}\BibitemShut {NoStop}%
\bibitem [{\citenamefont {Krauser}\ \emph {et~al.}(2014)\citenamefont
  {Krauser}, \citenamefont {Ebling}, \citenamefont {Fl\"aschner}, \citenamefont
  {Heinze}, \citenamefont {Sengstock}, \citenamefont {Lewenstein},
  \citenamefont {Eckardt},\ and\ \citenamefont {Becker}}]{Kra2014S}%
  \BibitemOpen
  \bibfield  {author} {\bibinfo {author} {\bibfnamefont {J.~S.}\ \bibnamefont
  {Krauser}}, \bibinfo {author} {\bibfnamefont {U.}~\bibnamefont {Ebling}},
  \bibinfo {author} {\bibfnamefont {N.}~\bibnamefont {Fl\"aschner}}, \bibinfo
  {author} {\bibfnamefont {J.}~\bibnamefont {Heinze}}, \bibinfo {author}
  {\bibfnamefont {K.}~\bibnamefont {Sengstock}}, \bibinfo {author}
  {\bibfnamefont {M.}~\bibnamefont {Lewenstein}}, \bibinfo {author}
  {\bibfnamefont {A.}~\bibnamefont {Eckardt}}, \ and\ \bibinfo {author}
  {\bibfnamefont {C.}~\bibnamefont {Becker}},\ }\href {\doibase
  10.1126/science.1244059} {\bibfield  {journal} {\bibinfo  {journal}
  {Science}\ }\textbf {\bibinfo {volume} {343}},\ \bibinfo {pages} {157}
  (\bibinfo {year} {2014})}\BibitemShut {NoStop}%
\bibitem [{\citenamefont {Bernier}\ \emph {et~al.}(2009)\citenamefont
  {Bernier}, \citenamefont {Kollath}, \citenamefont {Georges}, \citenamefont
  {De~Leo}, \citenamefont {Gerbier}, \citenamefont {Salomon},\ and\
  \citenamefont {K\"ohl}}]{BerPRA2009}%
  \BibitemOpen
  \bibfield  {author} {\bibinfo {author} {\bibfnamefont {J.-S.}\ \bibnamefont
  {Bernier}}, \bibinfo {author} {\bibfnamefont {C.}~\bibnamefont {Kollath}},
  \bibinfo {author} {\bibfnamefont {A.}~\bibnamefont {Georges}}, \bibinfo
  {author} {\bibfnamefont {L.}~\bibnamefont {De~Leo}}, \bibinfo {author}
  {\bibfnamefont {F.}~\bibnamefont {Gerbier}}, \bibinfo {author} {\bibfnamefont
  {C.}~\bibnamefont {Salomon}}, \ and\ \bibinfo {author} {\bibfnamefont
  {M.}~\bibnamefont {K\"ohl}},\ }\href {\doibase 10.1103/PhysRevA.79.061601}
  {\bibfield  {journal} {\bibinfo  {journal} {Phys. Rev. A}\ }\textbf {\bibinfo
  {volume} {79}},\ \bibinfo {pages} {061601} (\bibinfo {year}
  {2009})}\BibitemShut {NoStop}%
\bibitem [{\citenamefont {J\"ordens}\ \emph {et~al.}(2010)\citenamefont
  {J\"ordens}, \citenamefont {Tarruell}, \citenamefont {Greif}, \citenamefont
  {Uehlinger}, \citenamefont {Strohmaier}, \citenamefont {Moritz},
  \citenamefont {Esslinger}, \citenamefont {De~Leo}, \citenamefont {Kollath},
  \citenamefont {Georges}, \citenamefont {Scarola}, \citenamefont {Pollet},
  \citenamefont {Burovski}, \citenamefont {Kozik},\ and\ \citenamefont
  {Troyer}}]{Jor2010PRL}%
  \BibitemOpen
  \bibfield  {author} {\bibinfo {author} {\bibfnamefont {R.}~\bibnamefont
  {J\"ordens}}, \bibinfo {author} {\bibfnamefont {L.}~\bibnamefont {Tarruell}},
  \bibinfo {author} {\bibfnamefont {D.}~\bibnamefont {Greif}}, \bibinfo
  {author} {\bibfnamefont {T.}~\bibnamefont {Uehlinger}}, \bibinfo {author}
  {\bibfnamefont {N.}~\bibnamefont {Strohmaier}}, \bibinfo {author}
  {\bibfnamefont {H.}~\bibnamefont {Moritz}}, \bibinfo {author} {\bibfnamefont
  {T.}~\bibnamefont {Esslinger}}, \bibinfo {author} {\bibfnamefont
  {L.}~\bibnamefont {De~Leo}}, \bibinfo {author} {\bibfnamefont
  {C.}~\bibnamefont {Kollath}}, \bibinfo {author} {\bibfnamefont
  {A.}~\bibnamefont {Georges}}, \bibinfo {author} {\bibfnamefont
  {V.}~\bibnamefont {Scarola}}, \bibinfo {author} {\bibfnamefont
  {L.}~\bibnamefont {Pollet}}, \bibinfo {author} {\bibfnamefont
  {E.}~\bibnamefont {Burovski}}, \bibinfo {author} {\bibfnamefont
  {E.}~\bibnamefont {Kozik}}, \ and\ \bibinfo {author} {\bibfnamefont
  {M.}~\bibnamefont {Troyer}},\ }\href {\doibase
  10.1103/PhysRevLett.104.180401} {\bibfield  {journal} {\bibinfo  {journal}
  {Phys. Rev. Lett.}\ }\textbf {\bibinfo {volume} {104}},\ \bibinfo {pages}
  {180401} (\bibinfo {year} {2010})}\BibitemShut {NoStop}%
\bibitem [{\citenamefont {Lubasch}\ \emph {et~al.}(2011)\citenamefont
  {Lubasch}, \citenamefont {Murg}, \citenamefont {Schneider}, \citenamefont
  {Cirac},\ and\ \citenamefont {Ba\~nuls}}]{Lub2011PRL}%
  \BibitemOpen
  \bibfield  {author} {\bibinfo {author} {\bibfnamefont {M.}~\bibnamefont
  {Lubasch}}, \bibinfo {author} {\bibfnamefont {V.}~\bibnamefont {Murg}},
  \bibinfo {author} {\bibfnamefont {U.}~\bibnamefont {Schneider}}, \bibinfo
  {author} {\bibfnamefont {J.~I.}\ \bibnamefont {Cirac}}, \ and\ \bibinfo
  {author} {\bibfnamefont {M.-C.}\ \bibnamefont {Ba\~nuls}},\ }\href {\doibase
  10.1103/PhysRevLett.107.165301} {\bibfield  {journal} {\bibinfo  {journal}
  {Phys. Rev. Lett.}\ }\textbf {\bibinfo {volume} {107}},\ \bibinfo {pages}
  {165301} (\bibinfo {year} {2011})}\BibitemShut {NoStop}%
\bibitem [{\citenamefont {Hazzard}\ \emph {et~al.}(2012)\citenamefont
  {Hazzard}, \citenamefont {Gurarie}, \citenamefont {Hermele},\ and\
  \citenamefont {Rey}}]{Haz2012PRA}%
  \BibitemOpen
  \bibfield  {author} {\bibinfo {author} {\bibfnamefont {K.~R.~A.}\
  \bibnamefont {Hazzard}}, \bibinfo {author} {\bibfnamefont {V.}~\bibnamefont
  {Gurarie}}, \bibinfo {author} {\bibfnamefont {M.}~\bibnamefont {Hermele}}, \
  and\ \bibinfo {author} {\bibfnamefont {A.~M.}\ \bibnamefont {Rey}},\ }\href
  {\doibase 10.1103/PhysRevA.85.041604} {\bibfield  {journal} {\bibinfo
  {journal} {Phys. Rev. A}\ }\textbf {\bibinfo {volume} {85}},\ \bibinfo
  {pages} {041604} (\bibinfo {year} {2012})}\BibitemShut {NoStop}%
\bibitem [{\citenamefont {Greif}\ \emph {et~al.}(2013)\citenamefont {Greif},
  \citenamefont {Uehlinger}, \citenamefont {Jotzu}, \citenamefont {Tarruell},\
  and\ \citenamefont {Esslinger}}]{Gre2013Sci}%
  \BibitemOpen
  \bibfield  {author} {\bibinfo {author} {\bibfnamefont {D.}~\bibnamefont
  {Greif}}, \bibinfo {author} {\bibfnamefont {T.}~\bibnamefont {Uehlinger}},
  \bibinfo {author} {\bibfnamefont {G.}~\bibnamefont {Jotzu}}, \bibinfo
  {author} {\bibfnamefont {L.}~\bibnamefont {Tarruell}}, \ and\ \bibinfo
  {author} {\bibfnamefont {T.}~\bibnamefont {Esslinger}},\ }\href {\doibase
  10.1126/science.1236362} {\bibfield  {journal} {\bibinfo  {journal}
  {Science}\ }\textbf {\bibinfo {volume} {340}},\ \bibinfo {pages} {1307}
  (\bibinfo {year} {2013})}\BibitemShut {NoStop}%
\bibitem [{\citenamefont {Cherng}\ \emph {et~al.}(2007)\citenamefont {Cherng},
  \citenamefont {Refael},\ and\ \citenamefont {Demler}}]{Che2007PRL}%
  \BibitemOpen
  \bibfield  {author} {\bibinfo {author} {\bibfnamefont {R.~W.}\ \bibnamefont
  {Cherng}}, \bibinfo {author} {\bibfnamefont {G.}~\bibnamefont {Refael}}, \
  and\ \bibinfo {author} {\bibfnamefont {E.}~\bibnamefont {Demler}},\ }\href
  {\doibase 10.1103/PhysRevLett.99.130406} {\bibfield  {journal} {\bibinfo
  {journal} {Phys. Rev. Lett.}\ }\textbf {\bibinfo {volume} {99}},\ \bibinfo
  {pages} {130406} (\bibinfo {year} {2007})}\BibitemShut {NoStop}%
\bibitem [{\citenamefont {Rapp}\ \emph {et~al.}(2008)\citenamefont {Rapp},
  \citenamefont {Hofstetter},\ and\ \citenamefont {Zar\'and}}]{Rap2008PRB}%
  \BibitemOpen
  \bibfield  {author} {\bibinfo {author} {\bibfnamefont {A.}~\bibnamefont
  {Rapp}}, \bibinfo {author} {\bibfnamefont {W.}~\bibnamefont {Hofstetter}}, \
  and\ \bibinfo {author} {\bibfnamefont {G.}~\bibnamefont {Zar\'and}},\ }\href
  {\doibase 10.1103/PhysRevB.77.144520} {\bibfield  {journal} {\bibinfo
  {journal} {Phys. Rev. B}\ }\textbf {\bibinfo {volume} {77}},\ \bibinfo
  {pages} {144520} (\bibinfo {year} {2008})}\BibitemShut {NoStop}%
\bibitem [{\citenamefont {Titvinidze}\ \emph {et~al.}(2011)\citenamefont
  {Titvinidze}, \citenamefont {Privitera}, \citenamefont {Chang}, \citenamefont
  {Diehl}, \citenamefont {Baranov}, \citenamefont {Daley},\ and\ \citenamefont
  {Hofstetter}}]{Tit2011NJP}%
  \BibitemOpen
  \bibfield  {author} {\bibinfo {author} {\bibfnamefont {I.}~\bibnamefont
  {Titvinidze}}, \bibinfo {author} {\bibfnamefont {A.}~\bibnamefont
  {Privitera}}, \bibinfo {author} {\bibfnamefont {S.-Y.}\ \bibnamefont
  {Chang}}, \bibinfo {author} {\bibfnamefont {S.}~\bibnamefont {Diehl}},
  \bibinfo {author} {\bibfnamefont {M.~A.}\ \bibnamefont {Baranov}}, \bibinfo
  {author} {\bibfnamefont {A.}~\bibnamefont {Daley}}, \ and\ \bibinfo {author}
  {\bibfnamefont {W.}~\bibnamefont {Hofstetter}},\ }\href
  {http://stacks.iop.org/1367-2630/13/i=3/a=035013} {\bibfield  {journal}
  {\bibinfo  {journal} {New Journal of Physics}\ }\textbf {\bibinfo {volume}
  {13}},\ \bibinfo {pages} {035013} (\bibinfo {year} {2011})}\BibitemShut
  {NoStop}%
\bibitem [{\citenamefont {Kan\'asz-Nagy}\ and\ \citenamefont
  {Zar\'and}(2012)}]{Kan2012PRB}%
  \BibitemOpen
  \bibfield  {author} {\bibinfo {author} {\bibfnamefont {M.}~\bibnamefont
  {Kan\'asz-Nagy}}\ and\ \bibinfo {author} {\bibfnamefont {G.}~\bibnamefont
  {Zar\'and}},\ }\href {\doibase 10.1103/PhysRevB.86.064519} {\bibfield
  {journal} {\bibinfo  {journal} {Phys. Rev. B}\ }\textbf {\bibinfo {volume}
  {86}},\ \bibinfo {pages} {064519} (\bibinfo {year} {2012})}\BibitemShut
  {NoStop}%
\bibitem [{\citenamefont {Rapp}\ and\ \citenamefont
  {Rosch}(2011)}]{Rap2011PRA}%
  \BibitemOpen
  \bibfield  {author} {\bibinfo {author} {\bibfnamefont {A.}~\bibnamefont
  {Rapp}}\ and\ \bibinfo {author} {\bibfnamefont {A.}~\bibnamefont {Rosch}},\
  }\href {\doibase 10.1103/PhysRevA.83.053605} {\bibfield  {journal} {\bibinfo
  {journal} {Phys. Rev. A}\ }\textbf {\bibinfo {volume} {83}},\ \bibinfo
  {pages} {053605} (\bibinfo {year} {2011})}\BibitemShut {NoStop}%
\bibitem [{\citenamefont {Inaba}\ and\ \citenamefont
  {Suga}(2013)}]{Ina2013MPL}%
  \BibitemOpen
  \bibfield  {author} {\bibinfo {author} {\bibfnamefont {K.}~\bibnamefont
  {Inaba}}\ and\ \bibinfo {author} {\bibfnamefont {S.-i.}\ \bibnamefont
  {Suga}},\ }\href {\doibase 10.1142/S0217984913300081} {\bibfield  {journal}
  {\bibinfo  {journal} {Mod. Phys. Lett.}\ }\textbf {\bibinfo {volume} {B27}},\
  \bibinfo {pages} {1330008} (\bibinfo {year} {2013})}\BibitemShut {NoStop}%
\bibitem [{\citenamefont {T\'oth}\ \emph {et~al.}(2010)\citenamefont {T\'oth},
  \citenamefont {L\"auchli}, \citenamefont {Mila},\ and\ \citenamefont
  {Penc}}]{Tot2010PRL}%
  \BibitemOpen
  \bibfield  {author} {\bibinfo {author} {\bibfnamefont {T.~A.}\ \bibnamefont
  {T\'oth}}, \bibinfo {author} {\bibfnamefont {A.~M.}\ \bibnamefont
  {L\"auchli}}, \bibinfo {author} {\bibfnamefont {F.}~\bibnamefont {Mila}}, \
  and\ \bibinfo {author} {\bibfnamefont {K.}~\bibnamefont {Penc}},\ }\href
  {\doibase 10.1103/PhysRevLett.105.265301} {\bibfield  {journal} {\bibinfo
  {journal} {Phys. Rev. Lett.}\ }\textbf {\bibinfo {volume} {105}},\ \bibinfo
  {pages} {265301} (\bibinfo {year} {2010})}\BibitemShut {NoStop}%
\bibitem [{\citenamefont {Assaad}(2005)}]{FAs2005PRB}%
  \BibitemOpen
  \bibfield  {author} {\bibinfo {author} {\bibfnamefont {F.~F.}\ \bibnamefont
  {Assaad}},\ }\href {\doibase 10.1103/PhysRevB.71.075103} {\bibfield
  {journal} {\bibinfo  {journal} {Phys. Rev. B}\ }\textbf {\bibinfo {volume}
  {71}},\ \bibinfo {pages} {075103} (\bibinfo {year} {2005})}\BibitemShut
  {NoStop}%
\bibitem [{\citenamefont {Paramekanti}\ and\ \citenamefont
  {Marston}(2007)}]{Par2007JP}%
  \BibitemOpen
  \bibfield  {author} {\bibinfo {author} {\bibfnamefont {A.}~\bibnamefont
  {Paramekanti}}\ and\ \bibinfo {author} {\bibfnamefont {J.~B.}\ \bibnamefont
  {Marston}},\ }\href {http://stacks.iop.org/0953-8984/19/i=12/a=125215}
  {\bibfield  {journal} {\bibinfo  {journal} {J. Phys.: Condens. Matter}\
  }\textbf {\bibinfo {volume} {19}},\ \bibinfo {pages} {125215} (\bibinfo
  {year} {2007})}\BibitemShut {NoStop}%
\bibitem [{\citenamefont {Corboz}\ \emph {et~al.}(2011)\citenamefont {Corboz},
  \citenamefont {L\"auchli}, \citenamefont {Penc}, \citenamefont {Troyer},\
  and\ \citenamefont {Mila}}]{Cor2011PRL}%
  \BibitemOpen
  \bibfield  {author} {\bibinfo {author} {\bibfnamefont {P.}~\bibnamefont
  {Corboz}}, \bibinfo {author} {\bibfnamefont {A.~M.}\ \bibnamefont
  {L\"auchli}}, \bibinfo {author} {\bibfnamefont {K.}~\bibnamefont {Penc}},
  \bibinfo {author} {\bibfnamefont {M.}~\bibnamefont {Troyer}}, \ and\ \bibinfo
  {author} {\bibfnamefont {F.}~\bibnamefont {Mila}},\ }\href {\doibase
  10.1103/PhysRevLett.107.215301} {\bibfield  {journal} {\bibinfo  {journal}
  {Phys. Rev. Lett.}\ }\textbf {\bibinfo {volume} {107}},\ \bibinfo {pages}
  {215301} (\bibinfo {year} {2011})}\BibitemShut {NoStop}%
\bibitem [{\citenamefont {Cai}\ \emph {et~al.}(2013)\citenamefont {Cai},
  \citenamefont {Hung}, \citenamefont {Wang},\ and\ \citenamefont
  {Wu}}]{Cai2013PRB}%
  \BibitemOpen
  \bibfield  {author} {\bibinfo {author} {\bibfnamefont {Z.}~\bibnamefont
  {Cai}}, \bibinfo {author} {\bibfnamefont {H.-H.}\ \bibnamefont {Hung}},
  \bibinfo {author} {\bibfnamefont {L.}~\bibnamefont {Wang}}, \ and\ \bibinfo
  {author} {\bibfnamefont {C.}~\bibnamefont {Wu}},\ }\href {\doibase
  10.1103/PhysRevB.88.125108} {\bibfield  {journal} {\bibinfo  {journal} {Phys.
  Rev. B}\ }\textbf {\bibinfo {volume} {88}},\ \bibinfo {pages} {125108}
  (\bibinfo {year} {2013})}\BibitemShut {NoStop}%
\bibitem [{\citenamefont {Honerkamp}\ and\ \citenamefont
  {Hofstetter}(2004)}]{Hon2004PRL}%
  \BibitemOpen
  \bibfield  {author} {\bibinfo {author} {\bibfnamefont {C.}~\bibnamefont
  {Honerkamp}}\ and\ \bibinfo {author} {\bibfnamefont {W.}~\bibnamefont
  {Hofstetter}},\ }\href {\doibase 10.1103/PhysRevLett.92.170403} {\bibfield
  {journal} {\bibinfo  {journal} {Phys. Rev. Lett.}\ }\textbf {\bibinfo
  {volume} {92}},\ \bibinfo {pages} {170403} (\bibinfo {year}
  {2004})}\BibitemShut {NoStop}%
\bibitem [{\citenamefont {Hermele}\ \emph {et~al.}(2009)\citenamefont
  {Hermele}, \citenamefont {Gurarie},\ and\ \citenamefont {Rey}}]{Her2009PRL}%
  \BibitemOpen
  \bibfield  {author} {\bibinfo {author} {\bibfnamefont {M.}~\bibnamefont
  {Hermele}}, \bibinfo {author} {\bibfnamefont {V.}~\bibnamefont {Gurarie}}, \
  and\ \bibinfo {author} {\bibfnamefont {A.~M.}\ \bibnamefont {Rey}},\ }\href
  {\doibase 10.1103/PhysRevLett.103.135301} {\bibfield  {journal} {\bibinfo
  {journal} {Phys. Rev. Lett.}\ }\textbf {\bibinfo {volume} {103}},\ \bibinfo
  {pages} {135301} (\bibinfo {year} {2009})}\BibitemShut {NoStop}%
\bibitem [{\citenamefont {Gorshkov}\ \emph {et~al.}(2010)\citenamefont
  {Gorshkov}, \citenamefont {Hermele}, \citenamefont {Gurarie}, \citenamefont
  {Xu}, \citenamefont {Julienne}, \citenamefont {Ye}, \citenamefont {Zoller},
  \citenamefont {Demler}, \citenamefont {Lukin},\ and\ \citenamefont
  {Rey}}]{Gor2010Nat}%
  \BibitemOpen
  \bibfield  {author} {\bibinfo {author} {\bibfnamefont {A.~V.}\ \bibnamefont
  {Gorshkov}}, \bibinfo {author} {\bibfnamefont {M.}~\bibnamefont {Hermele}},
  \bibinfo {author} {\bibfnamefont {V.}~\bibnamefont {Gurarie}}, \bibinfo
  {author} {\bibfnamefont {C.}~\bibnamefont {Xu}}, \bibinfo {author}
  {\bibfnamefont {P.~S.}\ \bibnamefont {Julienne}}, \bibinfo {author}
  {\bibfnamefont {J.}~\bibnamefont {Ye}}, \bibinfo {author} {\bibfnamefont
  {P.}~\bibnamefont {Zoller}}, \bibinfo {author} {\bibfnamefont
  {E.}~\bibnamefont {Demler}}, \bibinfo {author} {\bibfnamefont {M.~D.}\
  \bibnamefont {Lukin}}, \ and\ \bibinfo {author} {\bibfnamefont {A.~M.}\
  \bibnamefont {Rey}},\ }\href {\doibase 10.1038/nphys1535} {\bibfield
  {journal} {\bibinfo  {journal} {Nat. Phys.}\ }\textbf {\bibinfo {volume}
  {6}},\ \bibinfo {pages} {289} (\bibinfo {year} {2010})}\BibitemShut {NoStop}%
\bibitem [{\citenamefont {Sinkovicz}\ \emph {et~al.}(2013)\citenamefont
  {Sinkovicz}, \citenamefont {Zamora}, \citenamefont {Szirmai}, \citenamefont
  {Lewenstein},\ and\ \citenamefont {Szirmai}}]{Sin2013PRA}%
  \BibitemOpen
  \bibfield  {author} {\bibinfo {author} {\bibfnamefont {P.}~\bibnamefont
  {Sinkovicz}}, \bibinfo {author} {\bibfnamefont {A.}~\bibnamefont {Zamora}},
  \bibinfo {author} {\bibfnamefont {E.}~\bibnamefont {Szirmai}}, \bibinfo
  {author} {\bibfnamefont {M.}~\bibnamefont {Lewenstein}}, \ and\ \bibinfo
  {author} {\bibfnamefont {G.}~\bibnamefont {Szirmai}},\ }\href {\doibase
  10.1103/PhysRevA.88.043619} {\bibfield  {journal} {\bibinfo  {journal} {Phys.
  Rev. A}\ }\textbf {\bibinfo {volume} {88}},\ \bibinfo {pages} {043619}
  (\bibinfo {year} {2013})}\BibitemShut {NoStop}%
\bibitem [{\citenamefont {MacDonald}\ \emph {et~al.}(1988)\citenamefont
  {MacDonald}, \citenamefont {Girvin},\ and\ \citenamefont
  {Yoshioka}}]{Mac1988PRB}%
  \BibitemOpen
  \bibfield  {author} {\bibinfo {author} {\bibfnamefont {A.~H.}\ \bibnamefont
  {MacDonald}}, \bibinfo {author} {\bibfnamefont {S.~M.}\ \bibnamefont
  {Girvin}}, \ and\ \bibinfo {author} {\bibfnamefont {D.}~\bibnamefont
  {Yoshioka}},\ }\href {\doibase 10.1103/PhysRevB.37.9753} {\bibfield
  {journal} {\bibinfo  {journal} {Phys. Rev. B}\ }\textbf {\bibinfo {volume}
  {37}},\ \bibinfo {pages} {9753} (\bibinfo {year} {1988})}\BibitemShut
  {NoStop}%
\bibitem [{\citenamefont {Georgi}(1999)}]{Georgi1999}%
  \BibitemOpen
  \bibfield  {author} {\bibinfo {author} {\bibfnamefont {H.}~\bibnamefont
  {Georgi}},\ }\href@noop {} {\emph {\bibinfo {title} {Lie Algebras in Particle
  Physics}}}\ (\bibinfo  {publisher} {Westview Press},\ \bibinfo {address}
  {Cambridge, MA},\ \bibinfo {year} {1999})\BibitemShut {NoStop}%
\bibitem [{\citenamefont {Georges}\ \emph {et~al.}(1996)\citenamefont
  {Georges}, \citenamefont {Kotliar}, \citenamefont {Krauth},\ and\
  \citenamefont {Rozenberg}}]{Geo1996RMP}%
  \BibitemOpen
  \bibfield  {author} {\bibinfo {author} {\bibfnamefont {A.}~\bibnamefont
  {Georges}}, \bibinfo {author} {\bibfnamefont {G.}~\bibnamefont {Kotliar}},
  \bibinfo {author} {\bibfnamefont {W.}~\bibnamefont {Krauth}}, \ and\ \bibinfo
  {author} {\bibfnamefont {M.~J.}\ \bibnamefont {Rozenberg}},\ }\href {\doibase
  10.1103/RevModPhys.68.13} {\bibfield  {journal} {\bibinfo  {journal} {Rev.
  Mod. Phys.}\ }\textbf {\bibinfo {volume} {68}},\ \bibinfo {pages} {13}
  (\bibinfo {year} {1996})}\BibitemShut {NoStop}%
\bibitem [{\citenamefont {Helmes}\ \emph {et~al.}(2008)\citenamefont {Helmes},
  \citenamefont {Costi},\ and\ \citenamefont {Rosch}}]{Hel2008PRL}%
  \BibitemOpen
  \bibfield  {author} {\bibinfo {author} {\bibfnamefont {R.~W.}\ \bibnamefont
  {Helmes}}, \bibinfo {author} {\bibfnamefont {T.~A.}\ \bibnamefont {Costi}}, \
  and\ \bibinfo {author} {\bibfnamefont {A.}~\bibnamefont {Rosch}},\ }\href
  {\doibase 10.1103/PhysRevLett.100.056403} {\bibfield  {journal} {\bibinfo
  {journal} {Phys. Rev. Lett.}\ }\textbf {\bibinfo {volume} {100}},\ \bibinfo
  {pages} {056403} (\bibinfo {year} {2008})}\BibitemShut {NoStop}%
\bibitem [{\citenamefont {Snoek}\ \emph {et~al.}(2008)\citenamefont {Snoek},
  \citenamefont {Titvinidze}, \citenamefont {T\"oke}, \citenamefont {Byczuk},\
  and\ \citenamefont {Hofstetter}}]{Sno2008NJP}%
  \BibitemOpen
  \bibfield  {author} {\bibinfo {author} {\bibfnamefont {M.}~\bibnamefont
  {Snoek}}, \bibinfo {author} {\bibfnamefont {I.}~\bibnamefont {Titvinidze}},
  \bibinfo {author} {\bibfnamefont {C.}~\bibnamefont {T\"oke}}, \bibinfo
  {author} {\bibfnamefont {K.}~\bibnamefont {Byczuk}}, \ and\ \bibinfo {author}
  {\bibfnamefont {W.}~\bibnamefont {Hofstetter}},\ }\href
  {http://stacks.iop.org/1367-2630/10/i=9/a=093008} {\bibfield  {journal}
  {\bibinfo  {journal} {New J. Phys.}\ }\textbf {\bibinfo {volume} {10}},\
  \bibinfo {pages} {093008} (\bibinfo {year} {2008})}\BibitemShut {NoStop}%
\bibitem [{\citenamefont {Wu}\ and\ \citenamefont {Zhang}(2005)}]{Wu2005PRB}%
  \BibitemOpen
  \bibfield  {author} {\bibinfo {author} {\bibfnamefont {C.}~\bibnamefont
  {Wu}}\ and\ \bibinfo {author} {\bibfnamefont {S.-C.}\ \bibnamefont {Zhang}},\
  }\href {\doibase 10.1103/PhysRevB.71.155115} {\bibfield  {journal} {\bibinfo
  {journal} {Phys. Rev. B}\ }\textbf {\bibinfo {volume} {71}},\ \bibinfo
  {pages} {155115} (\bibinfo {year} {2005})}\BibitemShut {NoStop}%
\bibitem [{\citenamefont {Gull}\ \emph {et~al.}(2011)\citenamefont {Gull},
  \citenamefont {Millis}, \citenamefont {Lichtenstein}, \citenamefont
  {Rubtsov}, \citenamefont {Troyer},\ and\ \citenamefont
  {Werner}}]{Gul2011RMP}%
  \BibitemOpen
  \bibfield  {author} {\bibinfo {author} {\bibfnamefont {E.}~\bibnamefont
  {Gull}}, \bibinfo {author} {\bibfnamefont {A.~J.}\ \bibnamefont {Millis}},
  \bibinfo {author} {\bibfnamefont {A.~I.}\ \bibnamefont {Lichtenstein}},
  \bibinfo {author} {\bibfnamefont {A.~N.}\ \bibnamefont {Rubtsov}}, \bibinfo
  {author} {\bibfnamefont {M.}~\bibnamefont {Troyer}}, \ and\ \bibinfo {author}
  {\bibfnamefont {P.}~\bibnamefont {Werner}},\ }\href {\doibase
  10.1103/RevModPhys.83.349} {\bibfield  {journal} {\bibinfo  {journal} {Rev.
  Mod. Phys.}\ }\textbf {\bibinfo {volume} {83}},\ \bibinfo {pages} {349}
  (\bibinfo {year} {2011})}\BibitemShut {NoStop}%
\bibitem [{\citenamefont {Buchhold}(2012)}]{Buchhold2012}%
  \BibitemOpen
  \bibfield  {author} {\bibinfo {author} {\bibfnamefont {M.}~\bibnamefont
  {Buchhold}},\ }\emph {\bibinfo {title} {Topological Phases of Interacting
  Fermions in Optical Lattices with Artificial Gauge Fields}},\ \href@noop {}
  {Master's thesis},\ \bibinfo  {school} {Goethe-Universit\"at Frankfurt}
  (\bibinfo {year} {2012})\BibitemShut {NoStop}%
\bibitem [{\citenamefont {Com\u{a}nac}(2007)}]{Comanac2007}%
  \BibitemOpen
  \bibfield  {author} {\bibinfo {author} {\bibfnamefont {A.-B.}\ \bibnamefont
  {Com\u{a}nac}},\ }\emph {\bibinfo {title} {Dynamical Mean-Field Theory of
  Correlated Electron Systems: New Algorithms and Applications to Local
  Observables}},\ \href@noop {} {Ph.D. thesis},\ \bibinfo  {school} {Columbia
  University} (\bibinfo {year} {2007})\BibitemShut {NoStop}%
\bibitem [{\citenamefont {Boehnke}\ \emph {et~al.}(2011)\citenamefont
  {Boehnke}, \citenamefont {Hafermann}, \citenamefont {Ferrero}, \citenamefont
  {Lechermann},\ and\ \citenamefont {Parcollet}}]{Boe2011PRB}%
  \BibitemOpen
  \bibfield  {author} {\bibinfo {author} {\bibfnamefont {L.}~\bibnamefont
  {Boehnke}}, \bibinfo {author} {\bibfnamefont {H.}~\bibnamefont {Hafermann}},
  \bibinfo {author} {\bibfnamefont {M.}~\bibnamefont {Ferrero}}, \bibinfo
  {author} {\bibfnamefont {F.}~\bibnamefont {Lechermann}}, \ and\ \bibinfo
  {author} {\bibfnamefont {O.}~\bibnamefont {Parcollet}},\ }\href {\doibase
  10.1103/PhysRevB.84.075145} {\bibfield  {journal} {\bibinfo  {journal} {Phys.
  Rev. B}\ }\textbf {\bibinfo {volume} {84}},\ \bibinfo {pages} {075145}
  (\bibinfo {year} {2011})}\BibitemShut {NoStop}%
\bibitem [{\citenamefont {Hafermann}\ \emph {et~al.}(2012)\citenamefont
  {Hafermann}, \citenamefont {Patton},\ and\ \citenamefont
  {Werner}}]{Haf2012PRB}%
  \BibitemOpen
  \bibfield  {author} {\bibinfo {author} {\bibfnamefont {H.}~\bibnamefont
  {Hafermann}}, \bibinfo {author} {\bibfnamefont {K.~R.}\ \bibnamefont
  {Patton}}, \ and\ \bibinfo {author} {\bibfnamefont {P.}~\bibnamefont
  {Werner}},\ }\href {\doibase 10.1103/PhysRevB.85.205106} {\bibfield
  {journal} {\bibinfo  {journal} {Phys. Rev. B}\ }\textbf {\bibinfo {volume}
  {85}},\ \bibinfo {pages} {205106} (\bibinfo {year} {2012})}\BibitemShut
  {NoStop}%
\bibitem [{\citenamefont {Bulla}\ \emph {et~al.}(1998)\citenamefont {Bulla},
  \citenamefont {Hewson},\ and\ \citenamefont {Pruschke}}]{Bulla1998JP}%
  \BibitemOpen
  \bibfield  {author} {\bibinfo {author} {\bibfnamefont {R.}~\bibnamefont
  {Bulla}}, \bibinfo {author} {\bibfnamefont {A.~C.}\ \bibnamefont {Hewson}}, \
  and\ \bibinfo {author} {\bibfnamefont {T.}~\bibnamefont {Pruschke}},\ }\href
  {http://stacks.iop.org/0953-8984/10/i=37/a=021} {\bibfield  {journal}
  {\bibinfo  {journal} {J. Phys.: Condens. Matter}\ }\textbf {\bibinfo {volume}
  {10}},\ \bibinfo {pages} {8365} (\bibinfo {year} {1998})}\BibitemShut
  {NoStop}%
\bibitem [{\citenamefont {Sotnikov}\ \emph {et~al.}(2012)\citenamefont
  {Sotnikov}, \citenamefont {Cocks},\ and\ \citenamefont
  {Hofstetter}}]{Sotnikov2012PRL}%
  \BibitemOpen
  \bibfield  {author} {\bibinfo {author} {\bibfnamefont {A.}~\bibnamefont
  {Sotnikov}}, \bibinfo {author} {\bibfnamefont {D.}~\bibnamefont {Cocks}}, \
  and\ \bibinfo {author} {\bibfnamefont {W.}~\bibnamefont {Hofstetter}},\
  }\href {\doibase 10.1103/PhysRevLett.109.065301} {\bibfield  {journal}
  {\bibinfo  {journal} {Phys. Rev. Lett.}\ }\textbf {\bibinfo {volume} {109}},\
  \bibinfo {pages} {065301} (\bibinfo {year} {2012})}\BibitemShut {NoStop}%
\bibitem [{\citenamefont {Sotnikov}\ \emph {et~al.}(2013)\citenamefont
  {Sotnikov}, \citenamefont {Snoek},\ and\ \citenamefont
  {Hofstetter}}]{Sotnikov2013PRA}%
  \BibitemOpen
  \bibfield  {author} {\bibinfo {author} {\bibfnamefont {A.}~\bibnamefont
  {Sotnikov}}, \bibinfo {author} {\bibfnamefont {M.}~\bibnamefont {Snoek}}, \
  and\ \bibinfo {author} {\bibfnamefont {W.}~\bibnamefont {Hofstetter}},\
  }\href {\doibase 10.1103/PhysRevA.87.053602} {\bibfield  {journal} {\bibinfo
  {journal} {Phys. Rev. A}\ }\textbf {\bibinfo {volume} {87}},\ \bibinfo
  {pages} {053602} (\bibinfo {year} {2013})}\BibitemShut {NoStop}%
\end{thebibliography}%

\end{document}